\documentclass[intlimits,twoside,a4paper]{article}

\usepackage{amsmath,amssymb}
\usepackage{graphicx}

\usepackage[T2A]{fontenc}
\usepackage[cp1251]{inputenc}
\usepackage[ukrainian,english]{babel}
\usepackage{color}

\def\ds{\displaystyle}
\newcommand\Pd{\tfrac{1}{2}}

\newcommand\vr{{\mathbf{r}}}
\newcommand\vk{{\mathbf{k}}}
\newcommand\vq{{\mathbf{q}}}
\newcommand\vp{{\mathbf{p}}}

\newcommand\be{\begin{equation}}
\newcommand\ee{\end{equation}}
\newcommand\bee{\begin{eqnarray}}
\newcommand\eee{\end{eqnarray}}

\usepackage[eqsecnum]{cmpj2}

\issue{2015}{18}{3}{33002}
\doinumber{10.5488/CMP.18.33002}

\title[Theory of many-boson system with deformed algebra]%
{Theory of a many-boson system with deformed Heisenberg algebra%
}
\author[I.O.~Vakarchuk,G.I.~Panochko]{I.O.~Vakarchuk \refaddr{label1},
        G.I.~Panochko\refaddr{label2}}
\addresses{
\addr{label1} Department for Theoretical Physics, Ivan Franko National University of Lviv, 12 Dragomanov St., 79005 Lviv, Ukraine
\addr{label2} College of Natural Sciences, Ivan Franko National University of Lviv, 107 Tarnavsky St., 79010 Lviv, Ukraine
}
\date{Received March 14, 2015, in final form May 20, 2015}

\begin{document}

\maketitle

\begin{abstract}
We propose to consider nonlinear fluctuations in the theory of liquid ${}^{4}$He deforming the commutation relations between the generalized coordinates and momenta.
 Generalized coordinates are coefficients of density fluctuations of Bose particles. The deformation parameter takes into account the effects of three- and four-particle correlations in the behavior of a system. This parameter is defined from the experimental values of the elementary excitation spectrum and the structure factor extrapolated to $T=0$~K.
The numerical estimation of the ground state energy and the Bose condensate fraction is made. The elementary excitation spectrum and the potential of interaction between the helium atoms are recovered.
\pacs 05.30.Jp, 02.40.Gh, 67.25.-k, 67.25.dt, 03.65.-w

\end{abstract}

\section{Introduction}

The method of collective variables is an effective approach to the study of an $N$-particle Bose system. In this method, independent variables are the Fourier coefficients
of the density fluctuations $\rho_{\vk}$ of the particle ($\vk$ is the wave vector).
This method was originally proposed by Bohm \cite{Bom,Bom1,Bom1_2,Bom1_3}.
It was suggested that from the whole set of an infinite number of variables $\rho_{\vk}$, only $DN$ variables are taken, where $D$ is the space dimensionality. To achieve this, the domain of the wave-vector absolute values was restricted by some~$k_{\mathrm{c}}$.

An obvious imperfection of this approach is an ambiguous choice of variables.
Bogoliubov and Zubarev \cite{Boh} proposed an approach where the set of values $\vk $ is not limited. However, the transition from Cartesian coordinates to collective variables $\rho_{\vk }$ can be provided
 by the weighting function. The hermitian Hamiltonian of the Bose liquid in the $\rho_{\vk }$ representation is written as a sum of Hamiltonians of an infinite number of non-interacting harmonic oscillators describing the oscillations
of the Bose-liquid density plus a contribution from anharmonicities of these oscillations.
Justifications of the feasibility of the method of collective variables are given in many articles using different approaches. Usually, the anharmonic contribution was considered using the perturbation operator
\cite{1,2,2_2,3,3_2,4,4_2,4_3,5,5_2,6,7,8,9,9_2}.
The respective results can be brought to the numerical calculations both for model systems and for strongly non-ideal  systems like liquid~${}^{4}\rm{He}$.

A different approach for the study of the Bose system arose following the introduction of
quantum spaces of minimal length.
The simplest deformation is that of Kempf being  quadratic in generalized momentum \cite{Kempf,Kempf1}.
The deformed commutation relations can be examined in multidimensional case.
 Such kind of deformation was applied to a wide variety of quantum mechanical problems, among which we distinguish the eigenvalue problem for $D$-dimensional isotropic harmonic oscillator \cite{Tkachuk1}, three-dimensional Dirac oscillator \cite{Tkachuk2}, (2+1)-dimensional Dirac equation in a constant magnetic
field \cite{Pedram} that were solved exactly. A composite system problem in the deformed space was also considered \cite{Buisseret, Tkachuk4}.
Some aspects of field theories were investigated in the deformed space, for example: the electromagnetic field \cite{Camacho,Camacho_2}, a photoelectric phenomenon \cite{Djak}, radiation and absorption of photons for deformed field \cite{10}, the Casimir effect for the deformed field \cite{11}.
An urgent problem is to find a possible generalization of arbitrary one-dimensional Heisenberg algebra with minimal length (or/and momentum) for multidimensional case \cite{Tkachuk3}. Kempf's deformation can be applied not only to the operators of positions and momenta but also to collective variables.
The idea to use a deformation of Poisson brackets to explore a Bose system firstly appeared in the article \cite{Monteiro1}.
In paper \cite{Monteiro2}, the Bose particles were represented as a set of $q$-deformed harmonic oscillators.
In paper \cite{Mizrahi}, there was used a representation of deformed creation and annihilation operators with a generalized four-parameter $q$-algebra. We note that the deformed creation and annihilation operators associate with the
non-linear $f$-oscillator operators. In paper \cite{Gavrilik}, two-parameter deformed Bose gas model  was proposed to find the correlation functions of the particle.
In a series of papers, the {thermodynamics of ideal Bose and
Fermi systems \cite{Lavagno,Lavagno_2,Fitjo}} was studied. In paper \cite{Zhang}, the phenomenon of the Bose-Einstein condensation of the relativistic ideal Bose gas
with deformed commutation relations for positions and momenta operators is investigated.

In our paper, we associate the difficulties connected with anharmonic contribution with
deformation of commutation relations between the positions and momenta.
Thus, we replace the multimode Hamiltonian of the Bose liquid in $\rho_k$ representation by a sum of
single-mode Hamiltonians. We assume that this model successfully describes the
properties of the Bose system when we make an appropriate choice of the
deformation parameter. In this approach, we treat the collective variables as
generalized coordinates. Thus, we attempt to take into account the anharmonic contribution
described in \cite{Vak} with deformed Poisson brackets quadratic in generalized momenta.
In this paper, we use the deformation function which is quadratic in generalized coordinates.
The correctness and efficiency of the proposed approach is shown for the analysis of Bose-systems with developed
anharmonisms of the density fluctuations.
Such a treatment of anharmonisms lowers the ground state energy if the deformation parameter is negative.
It is interesting to note that the energy levels are bounded for such values of the deformation parameter.
We make numerical estimations of the analytical results.

\section{Hamiltonian of Bose-liquid in the deformed space of collective variables}

Let us consider a system of $N$ spinless Bose particles of mass $m$  and  by the coordinates $\vr_{1}, \ldots, \vr_{N}$ which move in the $D$-dimension
space of the volume $V$. The Hamiltonian of the system reads
\be\label{21}
\hat{H}=\sum_{j=1}^{N}\frac{\hat{\vp}_{j}{}^{2}}{2m}+\sum_{1\leqslant
i<j\leqslant N}\Phi(|\vr_{i}-\vr_{j}|),\ee where the first term
is the kinetic energy operator, $\hat{\vp}_{j}=-\ri\hbar{\nabla}_{j}$
is the momentum operator of the $j$th particle; the second term is the potential energy consisting of a sum of the particle interaction potentials $\Phi(|\vr_{i}-\vr_{j}|)$.
We introduce the collective coordinates representation which takes the form:
\be\label{22} \rho_{\vk
}=\frac{1}{\sqrt{N}}\sum_{j=1}^{N}\re^{-\ri\vk \vr_{j}}, \qquad \vk
\neq0.\ee
The Hamiltonian of the Bose liquid (\ref{21}) can be rewritten in the following form \cite{Boh,1}:
\be\label{23}
\hat{H}=\sum_{\vk
\neq0}\frac{\hbar^{2}k^{2}}{2m}\left(-\frac{\partial^{2}}{\partial
\rho_{\vk }\partial \rho_{-\vk }}+\frac{1}{4}\rho_{\vk }\rho_{-\vk
}-\frac{1}{2}\right)+
\frac{N(N-1)}{2V}\nu_{0}+\frac{N}{2V}\sum_{\vk \neq0}\nu_{k}
(\rho_{\vk }\rho_{-\vk }-1)+\Delta\hat{H},
\ee
where
\be\label{2311}
\nu_{k}=\int\re^{-\ri\vk \mathbf{R}}\Phi(R)\rd\mathbf{R}
\ee
is the Fourier image of the interaction potential. The operator
$\Delta\hat{H}$ contains all anharmonic terms and in addition contains the term from the specific linear anharmonic by $\rho_{\vk }$ and quadratic by
$\partial/\partial\rho_{\vk }$:
 \bee\nonumber
\Delta\hat{H}&=&\sum_{\vk
\neq0}\sum_{\vk'
\neq0}\frac{\hbar^2(\vk\vk')}{2m\sqrt{N}}\rho_{\vk+\vk'}
\frac{\partial^2}{\partial\rho_{\vk}\partial\rho_{\vk'}}\\
&&{}+\sum_{n\geqslant 3}\frac{(-)^n}{4n(n-1)(\sqrt{N})^{n-2}}\mathop{\sum_{{\bf
k}_1\neq0} \ldots \sum_{{\bf k}_n \neq0}}\limits_{{\bf
k}_1+\ldots+ {\bf k}_n=0}\frac{\hbar^2}{2m}
(k_1^2+\dots +k_n^2)\rho_{\vk_1\dots  }\rho_{\vk_n }\,.
\eee

The introduced in (\ref{22}) variables $\rho_{\vk}$ are complex and can be represented as a linear combination of real variables:
\bee\nonumber
\rho_{\vk}&=&\rho_{\vk}^{\mathrm{c}}-\ri\rho_{\vk}^{\mathrm{s}}\,,\\
\rho_{\vk}^{\mathrm{c}}&=&\frac{1}{\sqrt{N}}\sum_{j=1}^{N}\cos
\vk\vr_{j}\,,\qquad
\rho_{\vk}^{\mathrm{s}}=\frac{1}{\sqrt{N}}\sum_{j=1}^{N}\sin
\vk\vr_{j}\,.\,\nonumber
\eee
Since they are the complex conjugate value  $\rho_{\vk}^{*}=\rho_{-\vk}$, that is
$\rho_{\vk}^{\mathrm{c}}=\rho_{-\vk}^{\mathrm{c}}$, $\rho_{\vk}^{\mathrm{s}}=-\rho_{-\vk}$, and
the variables with a particular index value $\vk$ equal to $(-\vk)$.
Thus, independent variables are $\rho_{\vk}$ when the wave vector $\vk$
takes up values to the half-space of the wave vectors.
Taking into account all the remarks we have made, let us consider the harmonic part of the Hamiltonian of the Bose-liquid (\ref{23}) as
 an infinite set of non-interacting harmonic oscillators with oscillation frequency
$\omega_{k}$:
\be\label{31}
\hat{H}=\sum_{\mu=c,s}\sum_{\vk\neq0}{}\!\!'\left(
\frac{\hat{P}^{2}_{\vk,\mu}}{2m_{k}}+\frac{m_{k}\omega_{k}^{2}\hat{Q}^{2}_{\vk,\mu}}{2}
\right)+
\frac{N(N-1)}{2V}\nu_{0}-\sum_{\vk\neq0}\left(\frac{\hbar^{2}k^{2}}{4m}+\frac{N}{2V}\nu_{k}
\right)\,,
\ee
where the generalized momentum operator $\hat{P}_{\vk,\mu}$ is conjugate to the generalized coordinate $\hat{Q}_{\vk,\mu}$.
In the coordinate $\rho_{\vk}$-representations
\be\label{311}
\hat{Q}_{\vk,c}=\rho_{\vk}^{\mathrm{c}}\,,\qquad
\hat{Q}_{\vk,s}=\rho_{\vk}^{\mathrm{s}}\,.
\ee
Explicit form of the generalized momentum operator is as follows:
\be\label{312}
\hat{P}_{\vk,\mu}=-\ri\hbar\frac{\partial}{\partial
\rho_{\vk,\mu }}\,.
\ee
The $\sum_{\vk\neq0}'$ means that the sum takes the values $\vk$ just from the haft-space of the domain.
Comparing Hamiltonians (\ref{31}) and (\ref{23}) we have
\be\label{32}
 m_{k}=\frac{2m}{k^{2}}\,,\qquad \omega_{k}=\frac{\hbar
k^{2}}{2m}\alpha_{k}\,,\qquad
\alpha_{k}=\sqrt{1+\frac{2N}{V}\nu_{k}\bigg
/\frac{\hbar^{2}k^{2}}{2m}}\,.
\ee
We note that the operator
$\hat{P}_{\vk,\mu}$ has the dimension of action $\hbar$, and
$\hat{Q}_{\vk,\mu}$ is dimensionless.

{The Hamiltonian of the system (\ref{23}) contains harmonic terms}. We suggest that the anharmonicity $\Delta\hat{H}$ can be taken
into account by deformation of the commutation relations between generalized coordinates and momenta:
\be\label{33}
\hat{Q}_{\vk,\mu}\hat{P}_{\vk,\mu}-\hat{P}_{\vk,\mu}\hat{Q}_{\vk,\mu}=
\ri\hbar\bigl(1+\beta_{k}\hat{Q}_{\vk,\mu}^{2}\bigr),
\ee
where the dimensionless deformation parameter $\beta_{k}$ depends on the absolute value of the wave-vector. The operators $\hat{Q}_{\vk,\mu}$ and $\hat{P}_{\vk',\mu'}$
with different index commute.
The deformation parameter for ${}^{4}{\rm He}$ can take negative values.
We do not expect a full description of the properties of the Bose liquid Hamiltonian
(\ref{31}) with the condition (\ref{33}) but assume that this model describes the behavior of the many-boson systems.

\section{Energy levels and wave functions of a Bose liquid}

We find solutions of the stationary Schr{\"o}dinger equation from harmonic oscillators in the deformed space with Hamiltonian (\ref{31}).
We suppose that the deformation is positive. For this purpose, we use the canonically conjugated operators $\hat{q}_{\vk,\mu}$, $\hat{p}_{\vk,\mu}$ with the standard Heisenberg algebra:
 \be\label{41}
\hat{P}_{\vk,\mu}=\hat{p}_{\vk,\mu}\,,\qquad
\hat{Q}_{\vk,\mu}=\frac{\tan {(}
\hat{q}_{\vk,\mu}\sqrt{\beta_{k}}{)}}{\sqrt{\beta_{k}}} \,.
\ee
To solve the eigenvalue problem we use a representation of the deformed operators
${\hat{P}}_{\vk,\mu}$ and $\hat{Q}_{\vk,\mu}$ (\ref{33}) which express them in terms of canonically
conjugate operators $\hat{q}_{\vk,\mu}$, $\hat{p}_{\vk,\mu}$. The representation we have
chosen reads:
\be \label{412}\hat{q}_{\vk,\mu}\hat{p}_{\vk,\mu}-\hat{p}_{\vk,\mu}\hat{q}_{\vk,\mu}=
\ri\hbar\,.\ee
Taking into account the representation (\ref{41}) we rewrite Hamiltonian (\ref{31}) in the form:
\bee\label{42}
\hat{H}=\sum_{\mu=c,s}\sum_{\vk\neq0}{}\!\!'\left[\frac{\hat{p}_{\vk,\mu}^{\,2}}{2m_{k}}
+\frac{m_{k}\omega_{k}^{2}}{2} \frac{\tan^{2}
{(}\hat{q}_{\vk,\mu}\sqrt{\beta_{k}}{)}}{\beta_{k}}
\right]+
\frac{N(N-1)}{2V}\nu_{0}-\sum_{\vk\neq0}\left(
\frac{\hbar^{2}k^{2}}{4m}+\frac{N}{2V}\nu_{k}\right)\,.
\eee
The eigenvalues and the wave functions of the harmonic oscillator with deformed Heisenberg algebra were calculated in \cite{1}.
Taking into
consideration our notations (\ref{32}) we find:
\bee\label{43}\nonumber
E_{\ldots,\, n_{\vk,c},\,
\ldots;\,\ldots,\,n_{\vk,s},\,\ldots}&=&\sum_{\mu=c,s}\sum_{\vk\neq0}{}\!\!'
\frac{\hbar^{2}k^{2}}{2m}\alpha_{k}\left[\bigl(n_{\vk,\mu}+\Pd\bigr)\sqrt{1+
\left(\frac{\beta_{k}}{2\alpha_{k}}\right)^{2}}\right.\\
&&\left.{}+\frac{\beta_{k}}{2\alpha_{k}}
\bigl(n_{\vk,\mu}^{2}+n_{\vk,\mu}+\Pd\bigr)
\vphantom{\sqrt{1+
\left(\frac{\beta_{k}}{2\alpha_{k}}\right)^{2}}}
\right]+\frac{N(N-1)}{2V}\nu_{0}-
\sum_{\vk\neq0}
\biggl(\frac{\hbar^{2}k^{2}}{4m}+\frac{N}{2V}\nu_{k}\biggr)\,,
\eee
here, quantum numbers
$n_{\vk,\mu}=0,1,2,\ldots$, $\mu=c,s$.
In the case of positive deformation parameter, the energy spectrum is infinite.
The wave functions in the coordinate representation
 $\hat{q}_{\vk,\mu}=q_{\vk,\mu}$,
$\hat{p}_{\vk,\mu}=- \ri\hbar\partial/\partial q_{\vk,\mu}$
can be written as follows:
\be\label{44}
\psi_{\ldots,\, n_{\vk,c},\,
\ldots;\,\ldots,\,n_{\vk,s},\,\ldots}(\ldots,\,q_{\vk,\mu},\,\ldots)=\prod_{\vk\neq0}{}\!\!'
\prod_{\mu=c,s}\psi_{n_{\vk,\mu}}(q_{\vk,\mu})\,.
\ee
The ground-state wave function ($n_{\vk,\mu}=0$) reads:
\be\nonumber
\psi_{0}(q_{\vk,\mu})=\beta_{k}^{1/4}
\sqrt{\frac{\Gamma(\nu+1)}{\Gamma(1/2)\Gamma(\nu+1/2)}}\cos^{\nu}q,
\ee
for $n_{\vk,\mu}\geqslant 1$, we have
\bee\label{45}\nonumber \psi_{n}(q_{\vk,\mu})&=&\beta^{1/4}
\sqrt{\frac{\Gamma(\nu+n+1)\Gamma(n+2\nu)}
{n!\Gamma(1/2)\Gamma(\nu+n+1/2)\Gamma(2\nu+2n)}}
\\&&{}\times\biggl(-\frac{\rd}{\rd q}+\nu\tan
q\biggr)\cdots\biggl(-\frac{\rd}{\rd q}+(\nu+n-1)\tan
q\biggr)\cos^{\nu+n}q\,,\eee
here,
\bee\label{46}
\nu&=&\frac{1}{2}+\frac{\alpha_{k}}{\beta_{k}}\sqrt{1+\biggl(\frac{\beta_{k}}{2\alpha_{k}}
\biggr)^{2}}\,,\\ q&=&q_{\vk,\mu}\sqrt{\beta_k},\quad
n=n_{\vk,\mu}\,.\nonumber\eee

The wave functions are orthonormalized: \be\label{47}
\int\limits^{\pi/(2\sqrt{\beta_{k}})}_{-\pi/(2\sqrt{\beta_{k}})}
\psi_{n'}(q_{\vk,\mu})\psi_{n}(q_{\vk,\mu})\,\rd q_{\vk,\mu}=\delta_{n',n}\,.\ee
The explicit form of the wave function of the first excited state is as follows:
\begin{equation}\label{48}\nonumber
\psi_{1}(q_{\vk,\mu})=\beta_{k}^{1/4}
\sqrt{\frac{2\Gamma(\nu+2)}{\Gamma(1/2)\Gamma(\nu+1/2)}}\cos^{\nu}q\,\sin q.
\end{equation}
The energy spectrum is quadratic in quantum numbers $n_{\vk,\mu}$, similarly to the theory of anharmonic oscillator
in the case it takes into account the terms proportional to $\sim\hbar^{2}$
\cite{1,12}.

Now we consider the case $\beta_k < 0$.
Then \be\label{49}
[\hat{Q}_{\vk,\mu},\hat{P}_{\vk,\mu}]=\ri\hbar\left(1-|\beta_k|\hat{Q}_{
\vk,\mu }^{2}\right),
\ee
and we impose the requirement
\be\label{4920}
\left(1-|\beta_{k}|\left\langle\hat{Q}_{\vk,\mu }^{2}\right\rangle\right)>0,
\ee
here, the brackets $\left\langle(\dots )\right\rangle$ denote an average:
\begin{equation}\label{493}
\left\langle(\dots )\right\rangle=\int \psi_{0}(q_{\vk,\mu})(\dots )\psi^{*}_{0}(q_{\vk,\mu})\rd q_{\vk,\mu}\,.
\end{equation}
The uncertainty relation obtained for the deformed algebra (\ref{49}) leads to the fact that minimal uncertainty for the momentum operator is equal to zero.
The new canonically conjugated variables (\ref{41}) and a structure of Hamiltonian (\ref{42}) with the changes
$\beta_{k} \rightarrow |\beta_{k}|$, $\tan
{(}\hat{q}_{\vk,\mu}\sqrt{\beta_{k}}{)}\rightarrow \tanh {(}
\hat{q}_{\vk,\mu}\sqrt{|\beta_{k}|}{)}$ will be the same. Using the factorization method \cite{1} we can write the wave functions of
the transformed Hamiltonian for $(k,\mu)$ modes:
\be\label{494}\nonumber \psi_{0}(q_{\vk,\mu})=|\beta_{k}|^{1/4}
\sqrt{\frac{\Gamma(\nu+1/2)}{\Gamma(1/2)\Gamma(\nu)}}\frac{1}{\cosh^\nu q}\,,
\ee
when $n\geqslant 1:$
\bee\label{491}
\psi_{n}(q_{\vk,\mu})&=&|\beta_{k}|^{1/4}
\sqrt{\frac{\Gamma(\nu-n+1/2)\Gamma(2\nu-2n+1)}
{n!\Gamma(1/2)\Gamma(\nu-n)\Gamma(2\nu-n+1)}}
\nonumber\\&&{}\times\left(-\frac{\rd}{\rd q}+\nu\tanh
q\right)\cdots\left(-\frac{\rd}{\rd q}+(\nu-n+1)\tanh
q\right)\cosh^{-\nu+n}q\,,
\eee
here,
\bee\label{492}
\nu&=&-\frac{1}{2}+\frac{\alpha_{k}}{|\beta_{k}|}\sqrt{1+\biggl(\frac{|\beta_{k}|}{2\alpha_{k}}
\biggr)^{2}}\,,\\ q&=&q_{\vk,\mu}\sqrt{|\beta_k|},\qquad
n=n_{\vk,\mu}\,. \nonumber \eee
In case of deformed algebra (\ref{49}), the quantum numbers are limited by the number $n<\nu$ and the energy levels are bounded from above. When $\nu\geqslant n$, the energy spectrum is continuous.
The wave function is also normalized. We note that the integration variable $q_{\vk,\mu}$ runs from
${\infty}$ up to ${-\infty}$
 \be\label{493}
\int\limits^{\infty}_{-\infty}
\psi_{n}(q_{\vk,\mu})\psi_{n}(q_{\vk,\mu})\,\rd q_{\vk,\mu}=1\,.
\ee
For example, we get for $n=1$:
\be\label{494}\nonumber
\psi_{1}(q_{\vk,\mu})=|\beta_{k}|^{1/4}
\sqrt{\frac{2\Gamma(\nu+1/2)}{\Gamma(1/2)\Gamma(\nu-1)}}\frac{\sinh
q}{\cosh^{\nu}q}\, \,.\ee
In the non-deformed case $(\beta_k=0)$, we obtain the energy levels of the Bose liquid in the main approximation \cite{1}.
When in equations~(\ref{45}) and (\ref{491}), for the wave functions, the quantity $\beta_{k}\to0$ and $\nu\to\infty$, we obtain:
$$\cos^{\nu}q\mathop{=}_{\beta_{k}\to0}\biggl(1-\frac{q_{\vk,\mu}^{2}\beta_{k}}{2}
+\cdots\biggr)^{\alpha_{k}/\beta_{k}}
\mathop{=}_{\beta_{k}\to0}\re^{-q^{2}_{\vk,\mu}\alpha_{k}/2}\,,$$

$$\cosh^{-\nu}q\mathop{=}_{|\beta_{k}|\to0}\biggl(1+\frac{q_{\vk,\mu}^{2}|\beta_{k}|}{2}
+\cdots\biggr)^{-\alpha_{k}/|\beta_{k}|}
\mathop{=}_{|\beta_{k}|\to0}\re^{-q^{2}_{\vk,\mu}\alpha_{k}/2}\,.$$
Taking into account the asymptotic formula for the gamma function
$\Gamma(\nu+a)\sim\sqrt{2\pi}\re^{-\nu}\nu^{\nu+a-1/2}$, when
$\nu\to\infty$, we obtain the wave functions of the harmonic oscillator:
 \be\label{52}
\psi_{n_{\vk,\mu}}(q_{\vk,\mu})=\biggl(\frac{\alpha_{k}}{\pi}\biggr)^{1/4}\frac{1}
{\sqrt{n_{\vk,\mu}!2^{n_{\vk,\mu}}}}\biggl(-\frac{\rd}{\rd\eta}+
\eta\biggr)^{n_{\vk,\mu}}\re^{-\eta^{2}/2}\,,\ee
here,
$\eta=q_{\vk,\mu}\sqrt{\alpha_{k}}, \quad
\alpha_{k}=m_k\omega_k/\hbar$.

\section{Ground-state energy}
If the quantum numbers in equation~(\ref{43}) are equal to zero,
 $n_{\vk,\mu}=0$, we obtain the ground-state energy $E_{0}$ which can be written after transformations in the form:
\bee\label{51}
E_{0}=\frac{N(N-1)}{2V}\nu_{0}-\sum_{\vk\neq0}\frac{\hbar^{2}k^{2}}{8m}(\alpha_{k}-1)^{2}
+\sum_{\vk\neq0}\frac{\hbar^{2}k^{2}}{4m}\alpha_{k}\left[\sqrt{1+\biggl(
\frac{\beta_{k}}{2\alpha_{k}}\biggr)^{2}}+\frac{\beta_{k}}{2\alpha_{k}}-1\right]\,.\eee
The first two terms recover the ground-state energy in the {Bogoliubov} approximation
\cite{13}. The last term in equation~(\ref{51}) takes into account the anharmonicity and always {leads to the negative
values for the parameter $\beta_k$}.
The lowering of the ground-state energy of the liquid ${}^{4}\rm{He}$ can by achieved by a direct consideration of
 the anharmonic operator $\Delta \hat{H}$ in the Hamiltonian (\ref{23}) from perturbation theory \cite{9,9_2}.
The Fourier coefficient $\nu_0$ in the limit $ k\rightarrow 0$  can be expressed via the speed of the first sound  for
liquid ${}^{4}{\rm He}$ (when $T=0$~ K, the speed is $c=238.2$\,m/s).
The definition of the speed of the first sound when $T=0$~K,
\be\label{52}\nonumber
 N\frac{\partial^2E_0}{\partial N^2}=mc^2,\ee
allows us to obtain the equation:
\be\label{53}
 mc^2=\frac{N}{V}\nu_0-\frac{1}{4N}\sum_{\vk\neq0}\frac{\hbar^2k^2}{4m}
 \frac{(\alpha_k^2-1)^2}{\alpha_k^3}\frac{1}{(1+(\beta_k/2\alpha_k)^2)^{3/2}}\,.\ee
Having taken into account the obtained equation (\ref{53}) we rewrite the ground-state energy in the thermodynamic limit:
\bee\label{511}
E_{0}&=&\frac{N m c^2}{2}-\frac{1}{4}\sum_{\vk\neq0}\frac{\hbar^{2}k^{2}}{2m}(\alpha_{k}-1)^{2}
+\frac{1}{2}\sum_{\vk\neq0}\frac{\hbar^{2}k^{2}}{2m}\alpha_{k}\left[\sqrt{1+\biggl(
\frac{\beta_{k}}{2\alpha_{k}}\biggr)^{2}}+\frac{\beta_{k}}{2\alpha_{k}}-1\right]\nonumber \\
&&{}+\frac{1}{16}\sum_{\vk\neq0}\frac{\hbar^{2}k^{2}}{2m}\frac{1}{\alpha_{k}}\biggl(\alpha_{k}
-\frac{1}{\alpha_{k}}\biggr)^{2}
\left[1+\left(\frac{\beta_{k}}{2\alpha_{k}}\right)^2\right]^{-3/2}\,.\eee
The found expressions can be used for the models with exact or perturbative solutions.
To verify the validity of the expression (\ref{511}), we might compare it with the results given by an exactly
solvable model or a model which allows perturbative consideration. Such a comparison brings about an
interesting question concerning the choice of deformation parameter $\beta_k$.

\section{Elementary excitation spectrum}

Let us find an expression for an elementary excitation given by the wave vector $\vq$.
Suppose that only the quantum
number $n_{\vq,c}=1$ and the other quantum numbers $n_{\vk,\mu}=0$ when $\vk\neq\vq$, $\mu\neq
c$.
From equation~(\ref{43}) we obtain:
$$E_{\ldots,\,0,\,n_{\vq,c}=1,\,0,\,\ldots;\,\ldots,\,0,\,\ldots}=
E_{\ldots,\,0,\,\ldots;\,\ldots,\,
0,\,n_{\vq,s}=1,\,0,\,\ldots;\,\ldots,\,0,\,\ldots}=E_{0}+E_{q}\,,$$
where the elementary excitation spectrum
\be\label{61}
E_{q}=\frac{\hbar^{2}q^{2}}{2m}\alpha_{q}\left[\sqrt{1+\biggl(\frac{\beta_{q}}{2\alpha_{q}}
\biggr)^{2}}+\frac{\beta_{q}}{\alpha_{q}}\right]\,.\ee
If $\beta_k=0$, the latter equation leads to the Bogoliubov's excitation spectrum \cite{13}:
\be\label{62}
E_{q}^{\mathrm{B}}=\frac{\hbar^{2}q^{2}}{2m}\alpha_{q}\,.\ee
Equation~(\ref{61}) is an exact solution of the  Schr{\"o}dinger equation with Hamiltonian (\ref{42}).
Here, we do not obtain the fading phenomenon of the elementary excitation because in Hamiltonian (\ref{42})
there are no terms that describe the dissipation and the decay of the elementary excitation.
The decay of the elementary excitations in the liquid ${}^{4}\rm{He}$ can be explained by the fact that the spectrum of Bose-liquid ends
at $k\simeq3.6$ \AA$^{-1}$ \cite{Woods}.

\section{Structure factor}
The structure factor of the system can be defined as an average of the square of the density fluctuations:
\be\label{71}
S_{k}=\langle|\rho_{\vk}|^{2}\rangle\,.\ee
Now, we calculate the average (\ref{71}) taking into consideration the fact that the collective variables are  generalized
ones and using the ground state wave functions (\ref{45}).
Since the collective variables are generalized coordinates $\hat{Q}_{\vk,\mu}$ from equation~(\ref{311}),
we find the average (\ref{71}) on the wave functions of the ground-state (\ref{45}):
\bee\label{72}
S_{k}&=&
\sum_{\mu=c,s}\langle\hat{Q}^{2}_{\vk,\mu}\rangle=\frac{1}{\beta_k}\sum_{\mu=c,s}\int\limits^{\pi/(2\sqrt{\beta_{k}})}_{-\pi/(2\sqrt{\beta_{k}})}{\psi_{0}^2
(q_{\vk,\mu})}\tan^2{\left(\sqrt{\beta_{k}}q_{\vk,\mu}\right)}\rd q_{\vk,\mu} \\\nonumber
&=& \ds \frac{2}{\beta_{k}} \frac{\Gamma(\nu+1)}{\Gamma(\nu+1/2)\Gamma(1/2)}\int\limits^{\pi/2}_{-\pi/2} \cos^{2\nu}q \tan^{2} q  \rd q\\
&=& \ds \frac{4}{\beta_{k}} \frac{\Gamma(\nu+1)}{\Gamma(\nu+1/2)\Gamma(1/2)}\int\limits^{\pi/2}_{-\pi/2} \sin^{2} q\cos^{2(\nu-1)}q   \rd q.\nonumber
\eee
This integral is reduced to the beta function:
\be\nonumber
S_{k}=\ds\frac{1}{\beta_{k}}\frac{\Gamma(\nu-1/2)}{\Gamma(\nu+1/2)}=\ds\frac{1}{\beta_{k}(\nu-1/2)}\,.
\ee
	The parameter $\nu$ should be taken from the relation  {\ref{46}} and after some transformations we obtain:
\be\label{73}
S_{k}=\ds\frac{1}{\alpha_{k}\sqrt{1+(\beta_{k}/2\alpha_{k})^{2}}}\,.\ee
When $\beta_{k}=0$, we recover the well-know result by {Bogoliubov} and Zubarev \cite{Boh},
\begin{equation}
S_{k}=1/\alpha_{k}\,.
\end{equation}
The elementary excitation spectrum (\ref{61}) reads:
\be\label{74}
E_{q}=\frac{\hbar^{2}q^{2}}{2m
S_q}+\frac{\hbar^{2}q^{2}}{2m}\beta_q\,.
\ee
The structure factor can be written as follows:
\[
S_{k}-1=\frac{2V}{N}\,\frac{\delta E_{0}}{\delta\nu_{k}}\,,
\]
here, $F$  is the free energy of the system, which is equal to the ground-state energy $E_{0}$ at $T=0$~K.
Having performed elementary calculations with equation~(\ref{51}) we obtain the result (\ref{73}).
Note that the structure factor $S_{k}$ is an analytic function of the deformation parameter $\beta_{k}$.
This fact can be seen when we calculated $S_{k}$ and make a parameter $\beta_k \rightarrow -|\beta_k|$ in the wave functions (\ref{491}):
\bee\label{732}
S_{k}&=&\sum_{\mu = c,s}\ds\frac{1}{|\beta_k|}\int_{-\infty}^{\infty}\psi_{0}^2
(q_{\vk,\mu})\tanh^2{\left(\sqrt{|\beta_k|} q_{\bf{k},\mu}\right)} \rd q_{\bf{k},\mu}\\
&=&
\ds \frac{2} {|\beta_k|}\frac{\Gamma(\nu+1/2)}{\Gamma(1/2)\Gamma(\nu)}
\int_{-\infty}^{\infty}\frac{\tanh^2 q}{\cosh^{2\nu} q} \rd q
=\ds \frac{4} {|\beta_k|}\frac{\Gamma(\nu+1/2)}{\Gamma(1/2)\Gamma(\nu)}
\int_{0}^{\infty}\frac{\sinh^2 q}{\cosh^{2\nu+2} q} \rd q .\nonumber
\eee
This integral is {reduced} to the beta function:
\be \nonumber
 S_{k}= \ds\frac{\Gamma(\nu+1/2)}{|\beta_k|\Gamma(\nu+3/2)}=\frac{1}{|\beta_k|(\nu+1/2)}\,,
\ee
here, the parameter $\nu$ is taken from the equation~(\ref{492}). Having performed some simplifications we obtain the structure factor from the equation~(\ref{73}).
{We note that condition (\ref{4920}) leading to the restrictions on the deformation parameter can be represented in the following form:}
\be\label{733}\ds \frac{|\beta_k|S_{k}}{2}<1.\ee

\section{Potential and kinetic energy}
Having used the structure factor we rewrite the ground-state energy of the many-body system (\ref{51}):
\be
\label{81}
E_0=\frac{N m c^2}{2}+\frac{1}{16}\sum_{\vk\neq0}\frac{\hbar^{2}k^{2}}{2m S_k}(S_k^2-1)^{2}
-\frac{1}{4}\sum_{\vk\neq0}\frac{\hbar^{2}k^{2}}{2m}\left(
1-\frac{1}{S_k}\right)^2
+\Delta E_0\,,\ee
here,
\be\label{82}
\Delta E_0=\frac{1}{16}\sum_{\vk\neq0}\frac{\hbar^{2}k^{2}}{2m S_k}\left[2(S_k^2-1)+
\left(\frac{\beta_k S_k}{2}\right)^2\right]\left(\frac{\beta_k S_k}{2}\right)^2+
\frac{1}{4}\sum_{\vk\neq0}\frac{\hbar^{2}k^{2}}{2m}\left[\beta_k+\left(\frac{\beta_k}{2}\right)^2\right].
\ee
The mean value for the potential energy can be calculated for the ground state
using equations~(\ref{53}), (\ref{73}). After some transformations, we arrive at the relation for the mean value of the potential energy:
\bee\label{75}
\langle\Phi\rangle&=&\frac{N(N-1)}{2V}\nu_{0}+\frac{N}{2V}\sum_{\vk\neq0}\nu_{k}
(\langle|\rho_{\vk}|^{2}\rangle-1)\nonumber\\&=&\frac{N(N-1)}{2V}\nu_{0}+
\sum_{\vk\neq0}\frac{\hbar^{2}k^{2}}{8m}(\alpha_{k}^{2}-1)
\left\{\frac{1}
{\alpha_{k}\sqrt{1+(\beta_{k}/2\alpha_{k})^{2}}}-1\right\}.
\eee
Let us rewrite the average potential energy in terms of $S_k$ and speed of the first sound:
\be
\label{77}
\langle\Phi\rangle=\frac{Nmc^2}{2}+\frac{1}{16}\sum_{\vk\neq0}\frac{\hbar^{2}k^{2}}{2m S_k}(S_k^2-1)^{2}-
\frac{1}{4}\sum_{\vk\neq0}\frac{\hbar^{2}k^{2}}{2m}(S_{k}-1)
\left(1-\frac{1}{S_{k}^2}\right)+\Delta \langle\Phi\rangle,\ee
here,
\be
 \label{771}
\Delta \langle\Phi\rangle=\frac{1}{32}\sum_{\vk\neq0}\frac{\hbar^{2}k^{2}}{2m}\beta_k^2(S_k-1)^2(S_k+2)+\frac{1}{16}\sum_{\vk\neq0}
\frac{\hbar^{2}k^{2}}{2m S_k}\ds\left(\frac{\beta_kS_k}{2}\right)^4.
\ee
The mean value of the kinetic energy is the average value of the operator (\ref{31}) in case $\nu_{k}=0$:
 \be\label{78} \langle
K\rangle=\sum_{\mu=c,s}{\sum_{\vk\neq0}}'\frac{\hbar^{2}k^{2}}{4m}\left[\frac{1}{\hbar^2}\langle
\hat{P}^2_{\vk,\mu}\rangle+\langle
\hat{Q}^2_{\vk,\mu}\rangle-1\right].
\ee
We calculate the first term in the square brackets on the ground state wave function from the equation~(\ref{45}):
\bee\label{82}
 \langle \hat{P}^2_{\vk,\mu}\rangle&=&-\hbar^2\frac{\Gamma(\nu+1)}{\Gamma(1/2)\Gamma(\nu+1/2)}\sqrt{\beta_k}
\int\limits^{\pi/(2\sqrt{\beta_k})}_{-\pi/(2\sqrt{\beta_k})} \cos^{\nu}q_{\bf{k},\mu} \frac{\partial^2}{\partial q^{2}_{\vk,\mu}}
\cos^{\nu}q_{\bf{k},\mu} \rd q_{\bf{k},\mu}\\
&=&\hbar^2\frac{\Gamma(\nu+1)}{\Gamma(1/2)\Gamma(\nu+1/2)}\sqrt{\beta_k}
\int\limits^{\pi/(2\sqrt{\beta_k})}_{-\pi/(2\sqrt{\beta_k})} \left|\frac{\partial}{\partial q_{\vk,\mu}}\cos^{\nu}q_{\bf{k},\mu} \right|^{2}\rd q_{\bf{k},\mu}\\ \nonumber
&=&2\beta_k \hbar^2\frac{\nu^2\Gamma(\nu+1)}{\Gamma(1/2)\Gamma(\nu+1/2)}\int\limits^{\pi/2}_{0} \sin^2 q\cos^{2\nu-2}q \rd q.\nonumber
\eee
After simple transformations we get:
\be\label{82} \langle \hat{P}^2_{\vk,\mu}\rangle=
 \frac{\hbar^2\beta_{k}\nu^2}{\nu-1/2}=\frac{\hbar^2}{2S_k} \left(1+\ds\frac{\beta_kS_k}{2}\right)^2\,,\ee
here, the expression for $\nu$ is taken from equation~(\ref{46}) and the structure factor from equation~(\ref{73}).

{The second term in equation~(\ref{78}) is the structure factor by definition. Finally, we can write:}
\be\label{99} \langle
K\rangle=\sum_{\vk\neq0}\frac{\hbar^{2}k^{2}}{8m} \left\{
\frac{\alpha_{k}[\sqrt{1+(\beta_{k}/2\alpha_{k})^{2}}+
\beta_{k}/2\alpha_{k}]^{2}}{\sqrt{1+(\beta_{k}/2\alpha_{k})^{2}}}
+\frac{1}{\alpha_{k}\sqrt{1+(\beta_{k}/2\alpha_{k})^{2}}}-2
 \right\}\,.\ee
{Taking into consideration equation~(\ref{73}) we obtain:}
 \be\label{79} \langle K\rangle =\ds\frac{1}{4}
 \sum_{\vk\neq0}\frac{\hbar^{2}k^{2}}{2m}\frac{\left(1-S_k\right)^{2}}{S_k}+\Delta \langle K\rangle,
\ee
 \be\label{791}
\Delta \langle K\rangle=\ds\frac{1}{4}\sum_{\vk\neq0}
\frac{\hbar^{2}k^{2}}{2m}\beta_k\Biggl(1+\frac{\beta_k S_k}{4}\Biggr).
\ee
This result for the kinetic energy can be obtained as a derivative from the ground state energy with respect to the mass:
 \be \label{80} \langle K\rangle = -m \frac{\partial E_0}{\partial m}\,.
 \ee
The sum of expressions (\ref{77}) and (\ref{79}) is the total energy (\ref{81}).

{Now we find the potential of the interaction between the Bose particles:}
\be \label{811} \Phi(r) = \ds\frac{1}{(2\pi)^3}\int \re^{\ri\vk\vr}\nu_k\ \rd \vk,
 \ee
here, $\nu_k$ is the Fourier coefficient of the interaction potential between Bose particles represented as a function of the structure
factor (\ref{73}):
\be \label{812} \nu_k = \ds\frac{V}{N}\frac{\hbar^2 k^2}{4m}\left(\ds\frac{1}{S_k^2}-1-\ds\frac{\beta_k^2}{4}\right).
 \ee
Thus, the interaction potential between the helium atoms reads:
\be\label{8011}
\Phi(r)=\ds\frac{1}{2\pi^2\rho r}\frac{\hbar^2}{2m}\int_0^\infty k^3\sin kr \left(\ds\frac{1}{S_k^2}-1\right)\ \rd k+\Delta \Phi(r),
\ee
here, the contribution is $\Delta \Phi(r)$ caused by the deformation of the commutation relations:
\be
\Delta \Phi(r)=-\ds\frac{1}{8\pi^2\rho r}\frac{\hbar^2}{2m}\int_0^\infty k^3\beta_k^2\sin kr \ \rd k.
\ee

\section{Momentum distribution}
To find the average number of atoms $N_{\vk}$ with {the} momentum $\hbar \vk$, $\vk\neq0$
 it is sufficient to calculate the variational derivative from the free energy of the system with respect to the free-particle
spectrum $\hbar^2k^2/2m$. We note that the free energy of the system coincides with the ground state energy when $T=0$~K [equation~(\ref{511}) by $n_{\vk,\mu}=0$].
After simple calculations we get:
\be
N_{\vk}\label{84}= \frac{1}{4}\left\{
\frac{\alpha_{k}[\sqrt{1+(\beta_{k}/2\alpha_{k})^{2}}
+\beta_{k}/2\alpha_{k}]^{2}}{\sqrt{1+(\beta_{k}/2\alpha_{k})^{2}}}
+\frac{1}{\alpha_{k}\sqrt{1+(\beta_{k}/2\alpha_{k})^{2}}}-2\right\},
\ee
or after some transformation
\be
N_{\vk}=\frac{1}{4S_k}\left[\left(1-S_k\right)^{2}+\beta_k S_k+{\Bigl(\frac{\beta_k S_k}{2}\Bigr)}^2\right].
\ee
The expression for the kinetic energy can be represented as a function of average numbers of particles:
\be\label{84a} \langle K \rangle= \sum_{\vk\neq 0}\frac{\hbar^2
k^2}{2m}N_{\vk}\,.\ee
{The average number of atoms $N_{\vk}$ with the momentum $\hbar \vk$ can be obtained from the expression for the mean value of the kinetic energy (\ref{84a}). To do so, we take into account the relation (\ref{99}).
After simple calculations, we obtain the expression for the average numbers of atoms $N_{\vk}$ which coincides with the expression   (\ref{84}).
We estimate the relative number of atoms in the case of their momenta being equal to zero (Bose condensate):}
\be\frac{N_{0}}{N}=1-\frac{1}{N}\sum_{\vk\neq0}N_{\vk}=1-\frac{1}{4N}\sum_{\vk\neq0}
\frac{\left(1-S_k\right)^{2}}{S_k}+\frac{\Delta N_0}{N}\,,
\ee
\be
\frac{\Delta N_0}{N}=-\frac{1}{4N}\sum_{\vk\neq0}
\beta_k\left(1+\ds\frac{\beta_k S_k}{4}\right).
\label{8.5}\ee
At intermediate calculation, we assume that ${\beta_k}$ is of positive value.
For numerical calculations of the deformation parameter it is negative.

\section{Deformation parameter. Numerical calculations}
For the numerical evaluation of the deformation parameter we proceed from the expression of the elementary excitation spectrum (\ref{74}). The value of the structure factor and excitation
spectrum are taken from experimental papers \cite{Doneli,Woods}:
\be\label{91}
\beta_{k} =\frac{E_k}{\hbar^2k^2/2m}-\frac{1}{S_k}\,.
\ee
We have the values of the deformation parameter for a limited range of wave vectors because the elementary excitation spectrum of the Bose liquid has the ultimate point of completion, and the experimental data of the structure factor are given up to 7.3~\AA$^{-1}$. However, as $k \to \infty$ the elementary excitation spectrum should be equal to the free-particle spectrum $E_k \to\hbar^2k^2/2m$ and the structure factor $S_{k}\to 1$.
\begin{figure}
\centerline{
\includegraphics[scale=0.28]{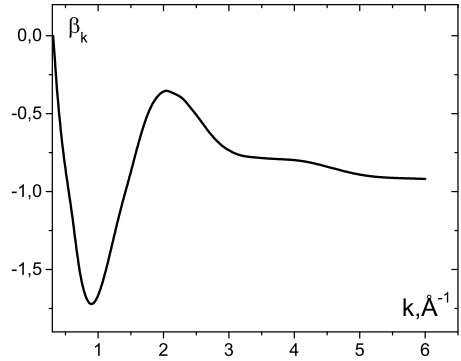}
}
\caption{Deformation parameter (\ref{91}). \label{fig0}}
\end{figure}
In figure~\ref{fig0}, the deformation parameter $\beta_k$ based on the experimental data for $S_{k}$ and $E_k$ is shown.
We shall model the deformation parameter by the function where a free parameter will be taken
from the extrapolated data for the structure factor of the Bose liquid at $T=0$~K \cite{vstup}:
\be\label{911}
\beta_{k} =-S_k|S_k-1|^3.
\ee
{The sign of the deformation parameter determines the behavior of the elementary excitation spectrum
we have received (6.5). Therefore, $\beta_k<0$ because the experimental data for the elementary excitation spectrum of the liquid helium (triangles in figure~\ref{fig6}, right panel) is lower than that of the theoretically calculated Feynman's spectrum  (circles in figure~\ref{fig6}, right panel).}

\begin{figure}[h!]
\includegraphics[scale=0.28]{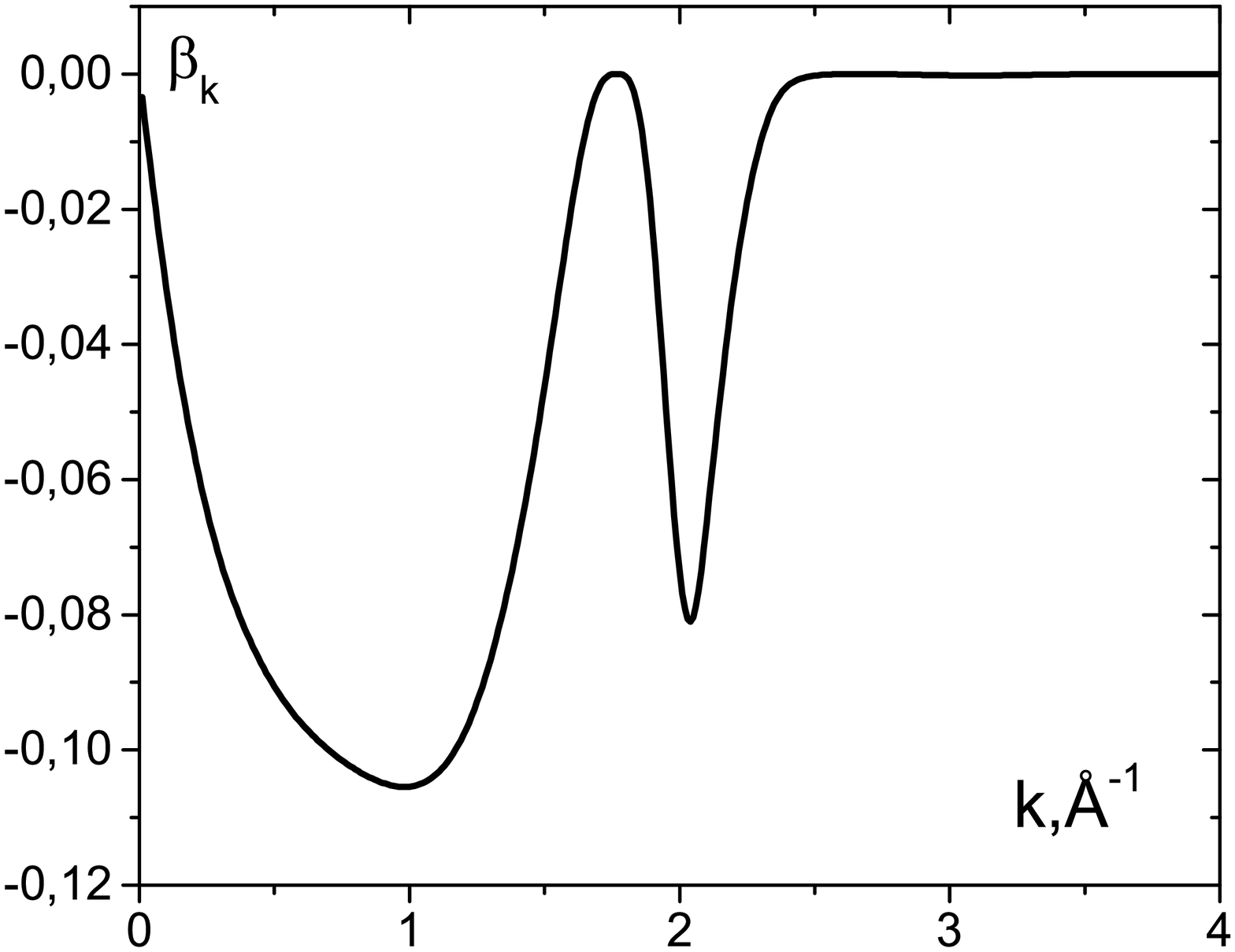}
\includegraphics[scale=0.28]{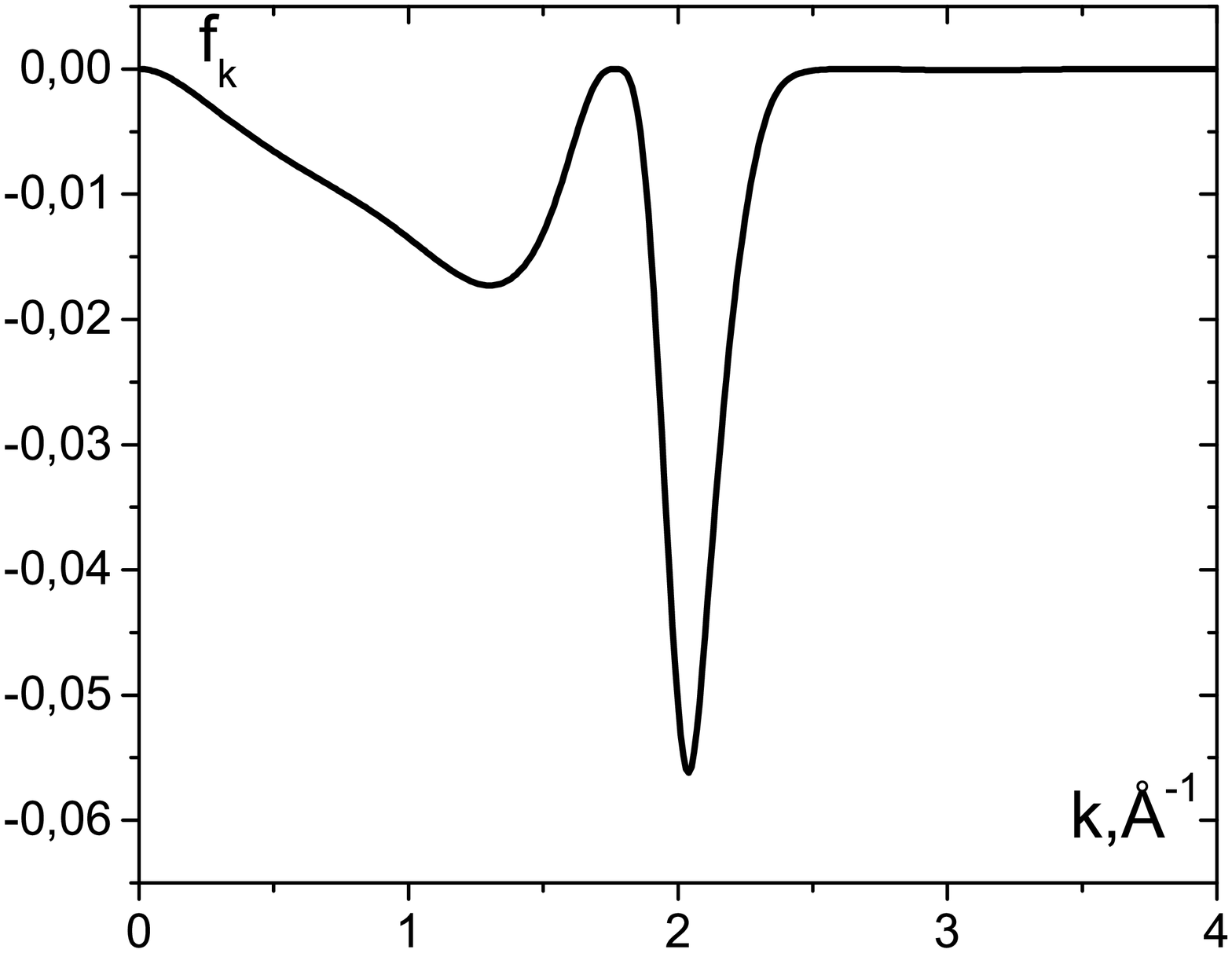}
\caption{(Left panel): Model of the deformation parameter (\ref{911});\label{fig1}
(Right panel): The function $f_k=\beta_kS_k/2$ as a condition to limit the deformation parameter (\ref{911}). \label{fig2}}
\end{figure}

\begin{figure}[!ht]
\includegraphics[scale=0.28]{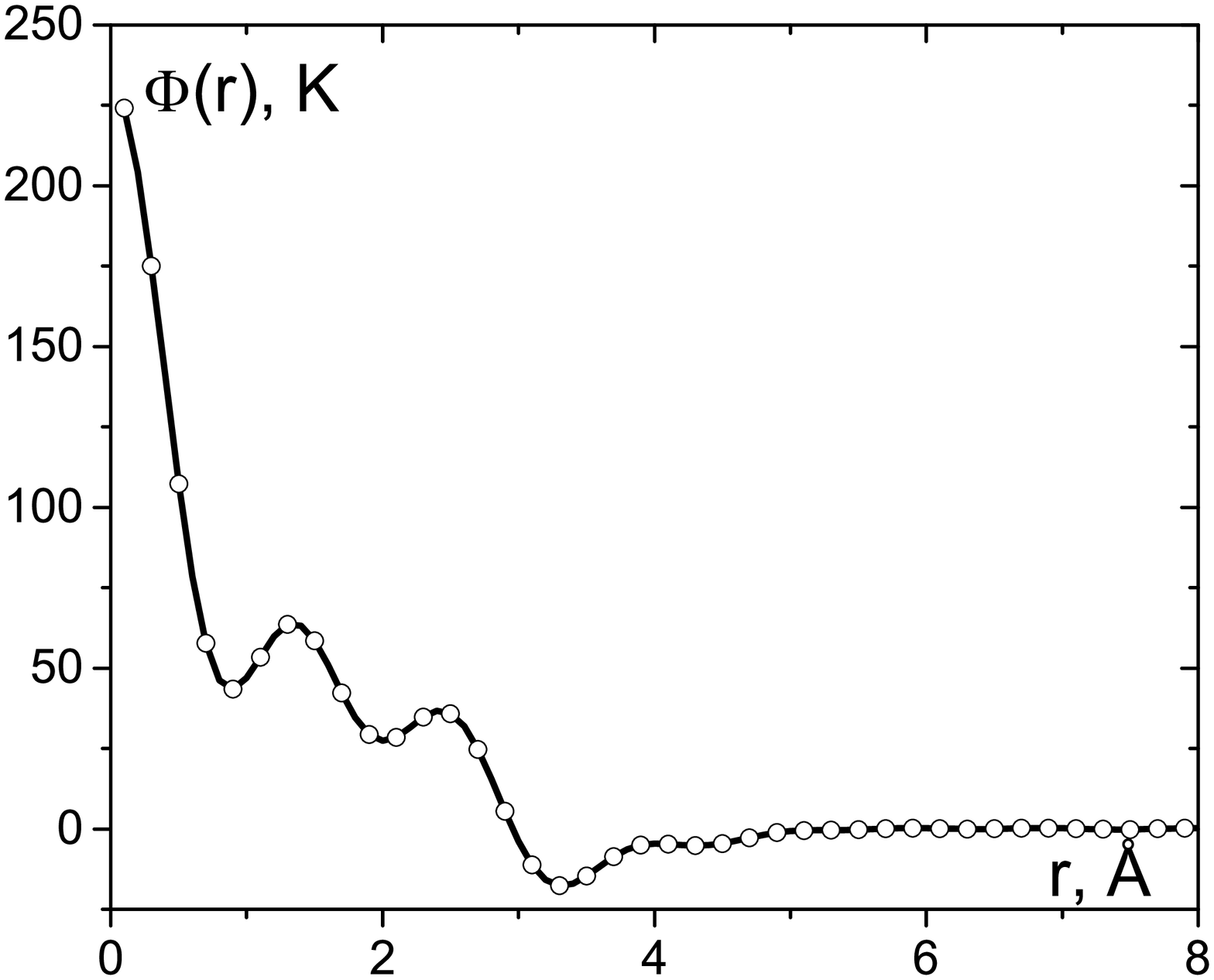}
\includegraphics[scale=0.28]{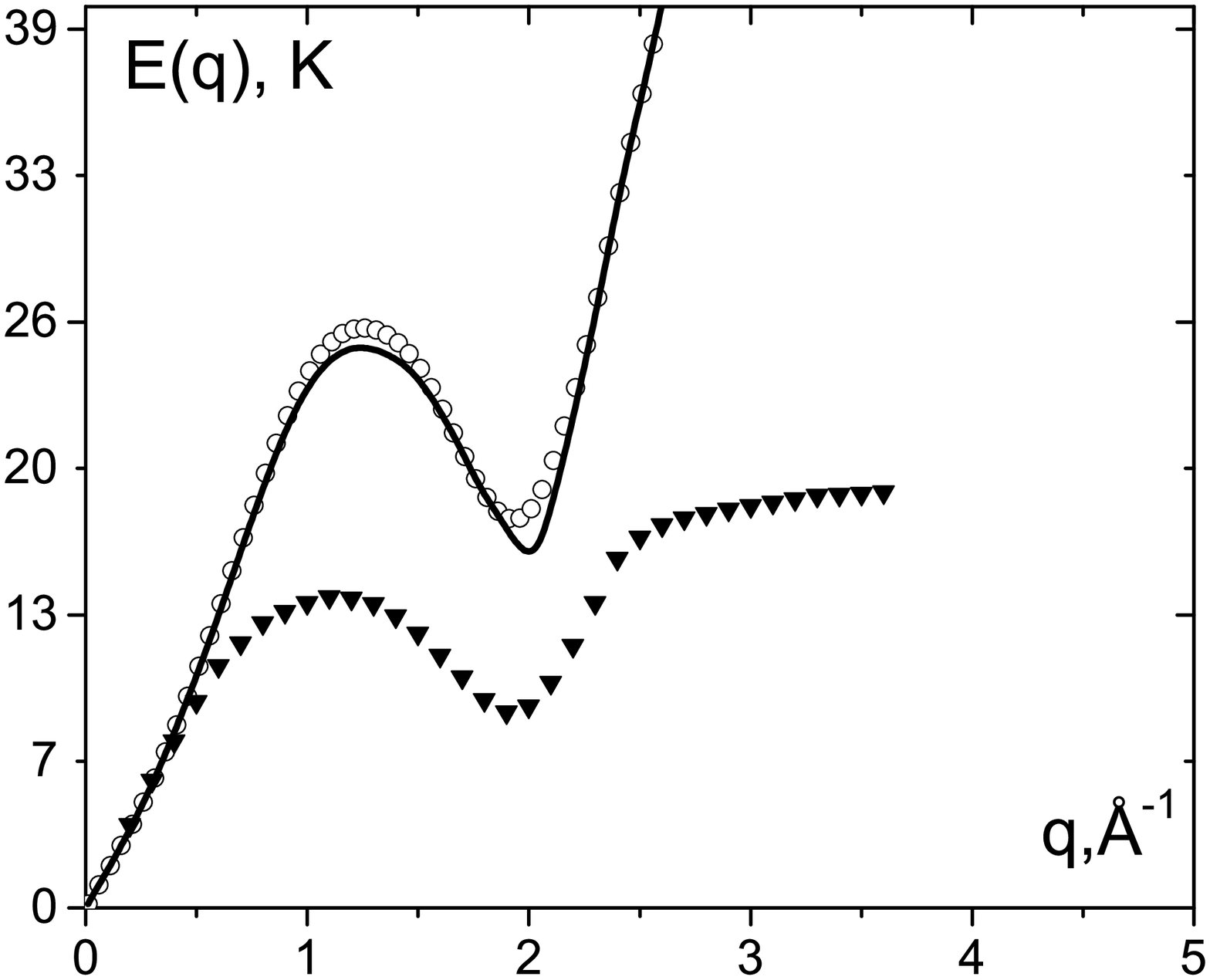}
\caption{(Left panel): The potential of the interaction between helium atoms. Circles correspond to the non-deformed case equation~(\ref {8011}) with $\Delta \Phi =0$. Solid line is the potential with models of the deformation parameter equation~(\ref{8011}) with $\Delta \Phi \neq 0$; \label{fig4}
(Right panel): The elementary excitation spectrum. Circles denote Feynman's spectrum; solid line is the spectrum of the deformed case; triangles correspond to the experimental data for the spectrum \cite{Woods}. \label{fig6}}
\end{figure}

The graph of the function (see equation~\ref{911}) is shown in figure~\ref{fig1} (left panel).
Note that this choice of the deformation parameter gives a correct behavior in the long-wavelength domain.
Function (\ref{911}) can be used to find the physical quantities in the limit of $T\to 0$.
{The form of the curve for $\beta_k$ (see figure~\ref{fig1}) should be such that the elementary excitation spectrum
(solid line in figure~\ref{fig6}) is reproduced for all the values of the wave vector. The point $k=1$~\AA$^{-1}$ corresponds to the typical maximum on the  curve of the elementary excitation spectrum (see figure~\ref{fig6}, right panel: triangles and circles). Thus, $\beta_k$ (see figure~\ref{fig1}) should have a clearly expressed minimum in this point.}
In figure~\ref{fig2} (right panel) the dependence reflecting limitations imposed by the deformation parameter (\ref{733}) is shown.

One can offer the model functions satisfying these considerations provided that contributions to the basic physical quantities of the system
reproduce the results in the post-RPA approximation in the theory of liquid helium-4.

With the deformation parameter from (\ref{911}), numerical calculations of the obtained quantities per particle
are made. In the thermodynamic limits ($V\to\infty$, $N\to\infty$,
$N/V={\rm const}$),
\be\nonumber
\sum_{\vk}\to \frac{V}{(2\pi)^3}\int \
\rd\vk, \ee
the contributions of the physical quantities connected with the deformed commutation relations are estimated.
We do not expect a complete agreement between the properties of the Bose liquid within the proposed method and the perturbative results. However, we obtained the results that are consistent with those of the perturbation theory.
The ground state energy with the deformation taken into consideration:
\be\label{93}\frac {E_0}{N}= \frac {E_0^{\mathrm{B}}}{N}+\frac {\Delta E_0}{N}\,,\qquad \frac {\Delta E_0}{N}=-1.89~{\rm K}\,.
\ee
The numerical value of the first term corresponds to the ground state energy in the Bogoliubov approximation, see \cite{vstup} (with $\rho = 0.0219$~\AA$^{-3}$ and speed of the first sound $c=238.2$~m/s): $E_0^{\mathrm{B}}/N= -5.31$~K.
Thus, the full energy per particle in the deformation case is $E_0/N= -7.2$~K.
The experimental result of the energy is $E_0/N= -7.13$~K.
Note that the linear corrections to the ground state energy over the deformation parameter give a leading contribution.
This choice of the deformation parameter (\ref{911}) offers an insignificant amendment to the value of the Bose condensate fraction obtained in the zeroth approximation:
\be\label{96}\frac {N_0}{N}= \frac {N_0^{\mathrm{B}}}{N}+\frac {\Delta N_0}{N}\,, \qquad \frac {\Delta N_0}{N}=0.11 .
\ee
In the zeroth approximation, $N_0^{\mathrm{B}}/N$ gives a wrong result: $N_0^{\mathrm{B}}/N=-0.31$.

{In our method, the anharmonic contributions from higher correlations in the Bose-liquid are taken into account within the single-mode approximation.
The Bose condensate fraction taking into account the correction $\Delta N_0/N$ in the deformed case [see equation~\eqref{8.5}] is calculated in
the approximation of the one sum over the wave vector. In this approximation, the value $N_0/N$ is negative.}

Nonlinear terms related to the deformation are quadratic in the expression for the interaction potential. This contribution vanishes, whereas the linear correction from the deformation parameter to the elementary excitation spectrum gives a significant contribution at $q\simeq2$~\AA$^{-1}$.
This correction does not contribute in the long-wavelength limit (figure~\ref{fig4}).
The behavior of the elementary excitation spectrum in the long-wavelength limit was solved in \cite{rov} of the method of two-time temperature Green's functions. At least a two-parametric deformation should be sought for to consider anharmonic {terms} $\Delta H$ in Hamiltonian (\ref{23}) in a proper way.
This issue will {be} a subject of our future studies.

\section*{Acknowledgements}
The authors appreciate the help in the preparation of this article of Ph.D.~Andrij Rovenchak and Ph.D.~Mykola Stetsko.

\ukrainianpart

\title{Теорія багатобозонної системи з деформованою алгеброю Гайзенберга}
\author{І.О.~Вакарчук\refaddr{label1}, Г.І.~Паночко\refaddr{label2}}
\addresses{
\addr{label1} Кафедра теоретичної фізики, Львівський національний університет імені Івана Франка, вул. Драгоманова 12, Львів, 79005, Україна
\addr{label2} Львівський національний університет імені Івана Франка, Природничий коледж, вул. Тарнавського 107, Львів, 79010, Україна
}
%
%
%

\makeukrtitle

\begin{abstract}
\tolerance=3000%
Запропоновано врахувати нелінійні флуктуації в теорії рідкого ${}^{4}\rm{He}$, деформуючи комутаційні співвідношення між узагальненими координатами та імпульсами. В якості узагальнених координат обрано коефіцієнти флуктуації густини бозе-частинок. Параметр деформації, що враховує вплив три- та чотиричастинкових кореляцій на поведінку бозе-систем, обрано виходячи з експериментальних значень для спектра елементарних збуджень та екстрапольованих експериментальних даних структурного фактора до температури $T=0$~K. З модельним параметром деформації
проведено чисельну оцінку енергії основного стану та кількості Бозе-конденсату, відтворено спектр елементарних збуджень та потенціал взаємодії між атомами ${}^{4}\rm{He}$.
\keywords деформована алгебра Гайзенберга, спектр елементарних збуджень рідкого ${}^{4}\rm{He}$, Бозе-конденсат

\end{abstract}


\begin{thebibliography}{99}
\bibitem{Bom1} Bohm~D., Pines~D., Phys. Rev., 1951, {\bf 82}, 625; \bibdoi{10.1103/PhysRev.82.625}.

\bibitem{Bom1_2} Bohm~D., Pines~D., Phys. Rev., 1952, {\bf 85}, 338; \bibdoi{10.1103/PhysRev.85.338}.

\bibitem{Bom1_3} Bohm~D., Pines~D., Phys. Rev., 1953, {\bf 92}, 609; \bibdoi{10.1103/PhysRev.92.609}.

\bibitem{Bom} Bohm~D., General Theory of Collective Coordinates, Mir, Moscow, 1964, 150~p. (in Russian).

\bibitem{Boh} Bogoliubov~N.N., Zubarev~D.N., Sov. Phys. JETP, 1955, {\bf 83}, 1
[Zh. Eksp. Fiz. Nizk. Temp., 1955, {\bf 28}, 129 (in Russian)].

\bibitem{1} Vakarchuk~I.O., Quantum Mechanics, Ivan Franko National University of Lviv, Lviv, 2012, 789--791; (in Ukrainian).

\bibitem{2} Hiroike~K., Prog. Theor. Phys., 1959, {\bf 21}, 327; \bibdoi{10.1143/PTP.21.327}.

\bibitem{2_2} Hiroike~K., Prog. Theor. Phys., 1975, {\bf54}, 308; \bibdoi{10.1143/PTP.54.308}.

\bibitem{3} Sunakawa~S., Yumasaki~S., Kebukawa~T., Prog. Theor. Phys., 1969,
{\bf 41}, 919, \bibdoi{10.1143/PTP.41.919}.


\bibitem{3_2} Sunakawa~S., Yumasaki~S., Kebukawa~T., Prog. Theor. Phys., 1975, {\bf 54}, 348 \bibdoi{10.1143/PTP.54.348}.

\bibitem{4} Vakarchuk~I.A., Yukhnovskii~I.R., Theor. Math. Phys., 1974, {\bf 18}, 63; \bibdoi{10.1007/BF01036928}.

\bibitem{4_2} Vakarchuk~I.A., Yukhnovskii~I.R., Theor. Math. Phys., 1979, {\bf 40}, 100; \doi{10.1007/BF01019246}.

\bibitem{4_3} Vakarchuk~I.A., Yukhnovskii~I.R., Theor. Math. Phys., 1980, {\bf 42}, 73; \bibdoi{10.1007/BF01019263}.

\bibitem{5} Grest~G.S., Rajagopal~A.K, Phys. Rev. A.,  1974, {\bf10}, 1395; \bibdoi{10.1103/PhysRevA.10.1395}.

\bibitem{5_2} Grest~G.S., Rajagopal~A.K, Phys. Rev. A.,  1974, {\bf10}, 1837; \bibdoi{10.1103/PhysRevA.10.1837}.

\bibitem{6} Sasaki~S., Matsuda~K., Prog. Theor. Phys., 1976, {\bf 56}, 375; \bibdoi{10.1143/PTP.56.375}.

\bibitem{7} Tserkovnikov~Yu.A., Theor. Math. Phys., 1976, {\bf 26}, 50; \bibdoi{10.1007/BF01038256}.

\bibitem{8} Vakarchuk~I.A., Glushak~P.A., Theor. Math. Phys., 1988, {\bf 75}, 399; \bibdoi{10.1007/BF01017174}.

\bibitem{9} Vakarchuk~I.A., Theor. Math. Phys., 1989, {\bf 80}, 983; \bibdoi{10.1007/BF01016193}.

\bibitem{9_2} Vakarchuk~I.A., Theor. Math. Phys., 1990, {\bf 82}, 308; \bibdoi{10.1007/BF01029225}.

\bibitem{Kempf} Kempf~A., J. Math. Phys., 1994, {\bf 35}, 4483; \bibdoi{10.1063/1.530798}.

\bibitem{Kempf1} Kempf~A., Mangano~G., Mann~R.B., Phys. Rev. D, 1995, {\bf 52}, 1108; \bibdoi{10.1103/PhysRevD.52.1108}.

\bibitem{Tkachuk1} Quesne~C., Tkachuk~V.M., J. Phys. A: Math. Gen., 2004, {\bf 37}, 10095; \bibdoi{10.1088/0305-4470/37/43/006}.

\bibitem{Tkachuk2} Quesne~C., Tkachuk~V.M., J. Phys. A: Math. Gen., 2005, {\bf 38}, 1747; \bibdoi{10.1088/0305-4470/38/8/011}.

\bibitem{Pedram} Pedram~P., Amirfakhrian~M., Shababi~H., Preprint \arxiv{1412.1425v2}, 2015.

\bibitem{Buisseret} Buisseret~F., Phys. Rev. A, 2010, {\bf 82}, 062102; \bibdoi{10.1103/PhysRevA.82.062102}.

\bibitem{Tkachuk4} Quesne~C., Tkachuk~V.M., Phys. Rev. A, 2010, {\bf81}, 012106; \bibdoi{10.1103/PhysRevA.81.012106}.

\bibitem{Tkachuk3} Nowicki~A., Tkachuk~V.M., J. Phys. A: Math. Theor., 2014, {\bf 47}, 025207; \bibdoi{10.1088/1751-8113/47/2/025207}.

\bibitem{Camacho} Camacho~A., Int. J. Mod. Phys. D, 2003, {\bf 12}, 1687; \bibdoi{10.1142/S0218271803004341}.

\bibitem{Camacho_2} Camacho~A., Gen. Relat. Gravit., 2003, {\bf 35}, 1153, \bibdoi{10.1023/A:1024437522212}.

\bibitem{Djak} Vakarchuk~I.O., Diakiv~Iu.M., J. Phys. Stud., 2014, {\bf 18}, 2004--1 (in Ukrainian).

\bibitem{10} Vakarchuk~I.O., Condens. Matter Phys., 2008, {\bf 11}, 409; \bibdoi{10.5488/CMP.11.3.409}.

\bibitem{11} Vakarchuk~I.O., J. Phys. A: Math. Theor., 2008, {\bf 41}, 185402; \bibdoi{10.1088/1751-8113/41/18/185402}.

\bibitem{Monteiro1} Monteiro~M.R., Rodrigues~L.M.C.S.,  Wulck~S., Phys. Rev. Lett., 1996, {\bf 76}, 1098; \bibdoi{10.1103/PhysRevLett.76.1098}.

\bibitem{Monteiro2} Monteiro~M.R., Rodrigues~L.M.C.S., Physica A, 1998, {\bf 259}, 245; \bibdoi{10.1016/S0378-4371(97)00633-X}.

\bibitem{Mizrahi} Mizrahi~S.S., Camargo Lima~J.P., Dodonov~V.V., J. Phys. A: Math. Gen., 2004, {\bf 37}, 11, 3707;
\bibdoi{10.1088/0305-4470/37/11/012}.

\bibitem{Gavrilik}  Gavrilik~A.M., Mishchenko~Yu.M., Nucl. Phys. B, 2015, {\bf 891}, 466; \bibdoi{10.1016/j.nuclphysb.2014.12.017}.

\bibitem{Lavagno} Lavagno~A., Swamy~P.N., Int. J. Mod. Phys. B, 2009, {\bf 23},  235; \bibdoi{10.1142/S0217979209049723}.

\bibitem{Lavagno_2} Lavagno~A., Swamy~P.N., Found. Phys., 2010, {\bf 40}, No.~7, 814; \bibdoi{10.1007/s10701-009-9363-0}.

\bibitem{Fitjo} Fityo~T.V., Phys. Lett. A, 2008, {\bf 372}, No.~37, 5872; \bibdoi{10.1016/j.physleta.2008.07.047}.

\bibitem{Zhang} Xiu-Ming~Z., Chi~T., Chinese Phys. Lett., 2015, {\bf 32}, 010303; \bibdoi{10.1088/0256-307X/32/1/010303}.

\bibitem{Vak}Vakarchuk~I.O., Ukr. J. Phys., 2010, {\bf 55}, No. 1, 36--43.

\bibitem{12} Landau~L.D., Lifshiz~E.M., Quantum Mechanics, Nauka, Moscow, 1974 (in Russian).

\bibitem{13} Bogoliubov~N., J. Phys. (USSR), 1947, {\bf 9}, 23.

\bibitem{Doneli} Donnelly~R.J., Donnelly~J. A., Hills~R.N., J. Low Temp. Phys., 1981, {\bf 44},
471; \bibdoi{10.1007/BF00117839}.

\bibitem{Woods} Cowley~R.A., Woods~A.D.B., Can. J. Phys., 1971, {\bf 49}, 177; \bibdoi{10.1139/p71-021}.

\bibitem{vstup} Vakarchuk~I.O., Introduction into the Many-Body Problem,  Ivan Franko National University of Lviv, Lviv, 1999 (in Ukrainian).

\bibitem{rov} Rovenchak~A., Z. Naturforsch. A, 2015, {\bf 70}, 73; \bibdoi{10.1515/zna-2014-0211}.

\end{thebibliography}
\end{document}